\documentclass[12pt]{article}
\usepackage{epsfig}
\usepackage{graphicx,psfrag,epsf}
\usepackage{amsbsy}
\usepackage{amsfonts}
\usepackage{amssymb}
\usepackage{amsmath, amsthm}
\usepackage{natbib}
\usepackage{caption}
\usepackage{setspace}
\usepackage{bm}
\usepackage{float}
\usepackage[usenames,dvipsnames]{color}
\usepackage{multirow}
\usepackage[T1]{fontenc}
\usepackage{lmodern}
\usepackage{amsmath}
\usepackage[figuresright]{rotating}
\usepackage{enumerate}
\usepackage{url}
\usepackage{bm}
\usepackage{xr}
\usepackage{tikz}
\usepackage{algorithm}
\usepackage[noend]{algpseudocode}
\usepackage{bbm}

\newcommand{\blind}{1}

\newcommand{\bbeta}{\boldsymbol{\beta}} 
\newcommand{\balpha}{\boldsymbol{\alpha}} 
\newcommand{\bmu}{\boldsymbol{\mu}}

\newcommand{\bth}{\boldsymbol{\theta}}    

\newcommand{\dd}{\, \mathrm{d}}

\begin{document}

\def\spacingset#1{\renewcommand{\baselinestretch}%
{#1}\small\normalsize} \spacingset{1}

\if1\blind
{
  \title{\bf Inferring Brain Signals Synchronicity from a Sample of EEG Readings}
  \author{Qian Li$^{1}$\thanks{This work was supported by the grant R01 GM111378-01A1 (DS, DT, CS) from the National Institute of General Medical Sciences.}\hspace{.2cm}\\
    Damla \c{S}ent\"{u}rk$^{1,2}$,\\
    Catherine A. Sugar$^{1,2,3}$,\\
    Shafali Jeste$^{3}$,\\
    Charlotte DiStefano$^{3}$,\\
    Joel Frohlich$^{3}$,\\
    and\\
    Donatello Telesca$^{1,*}$.\\
     $^1$ Department of Biostatistics, University of California, Los Angeles\\
     $^2$ Department of Statistics, University of California, Los Angeles\\
     $^3$ Department of Psychiatry and Biobehavioral Sciences,\\ University of California, Los Angeles\\
     $^*$ {\it email: dtelesca@ucla.edu}\\
    }
  \maketitle
} \fi

\if0\blind
{
  \bigskip
  \bigskip
  \bigskip
  \begin{center}
    {\LARGE\bf  Inferring Brain Signals Synchronicity from a Sample of EEG Readings}
\end{center}
  \medskip
} \fi

\newpage
\begin{abstract} %To be updated upon finishing
\noindent Inferring patterns of synchronous brain activity from a heterogeneous sample of electroencephalograms (EEG) is scientifically and methodologically challenging. While it is intuitively and statistically appealing to rely on readings from more than one individual in order to highlight recurrent patterns of brain activation, pooling information across subjects presents non-trivial methodological problems. We discuss some of the scientific issues associated with the understanding of synchronized neuronal activity and propose a methodological framework for statistical inference from a sample of EEG readings. Our work builds on classical contributions in time-series, clustering and functional data analysis, in an effort to reframe a challenging inferential problem in the context of familiar analytical techniques. Some attention is paid to computational issues, with a proposal based on the combination of machine learning and Bayesian techniques.  
\end{abstract}

\noindent%
\vfill

\newpage
\spacingset{1.45} % DON'T change the spacing!

%%% INTRO -------------------------------------------------------------------------------

\section{Introduction} 
\label{sec:eeg}

Functional neuroimaging technologies, including MRI, PET, MEG, and EEG,  aim to measure different aspects of brain function as they relate to specific mental processes.   This article focuses on the analysis of Electroencephalography (EEG) data in the context of neuropsychology studies.  EEG is a well-established noninvasive method for measuring spontaneous and event-related electrical activity across brain regions. The technology captures voltage fluctuation as signals, which reflect the distributed neuronal activities being projected on a cortical patch on which an EEG sensor is placed (\citealt{Teplan:2002}). The general aim of an EEG study is often the identification of neural function and cognitive states. Diverse biomedical applications include epilepsy, sleep disorders, multiple sclerosis, brain tumors, lesions, schizophrenia, and mood disorders (\citealt{Teplan:2002}).

Typical analyses in EEG studies focus primarily on inferring group differences in regions of interest. Such differences are assessed both in the frequency domain, by means of an amplified Signal-to-Noise Ratio (SNR) (\citealt{Laufs:2003}), and, in the case of  studies involving external stimuli,  in the time domain, by means of averaging and smoothing over repeated applications of the stimuli (\citealt{Hasenstab:2015}).

Beyond differential activation of brain regions, mounting evidence is building a case for the deeper understanding of neural interactions (\citealt{DiMartino:2014}, \citealt{Craddock:2013}). In this setting, magnetic resonance imaging has become an established workhorse for the mapping and annotation of the human connectome at the macro-scale.  The key to the success of MRI technologies as a preferred measurement tool in functional connectivity studies lies in their ability to produce measurements at high spatial resolution. This ability comes, however, at the cost of low time resolution, and perhaps most importantly, at the cost of severe hardware limitations, intended as the need to rely on expensive and bulky MRI scanners, which make MRI studies hard to design in a logistically and financially feasible fashion. 

On the other end,  EEG is thought to provide reliable measurements of neuronal activity only for the brain cortical regions, with low spatial resolution and often low SNR. However, compared to other imaging techniques, EEG has the advantage of relying on less bulky hardware and is associated with robust and extremely non-invasive imaging protocols, making the technology readily available for implementation and adaptation to a variety of scientific investigations.  

Recently , \cite{euan2015spectral}  suggested exploiting EEG's excellent temporal resolution by defining the concept of spectral synchronicity. In particular, a pair of EEG signals are considered spectrally synchronized if they are both dominated by similar frequency oscillations.  This idea formalizes the concept of coordinated neuronal activity and reflects recent empirical evidence, which suggests that differential patterns of coordinated neuronal activity may be associated with a range of neuropsychiatric and neurological processes, including memory formation (\citealt{Fell:2011}) and mental disorders (\citealt{Broyd:2008}).  

From a statistical perspective, multi-subject studies of functional connectivity still pose substantial methodological challenges. Ideally, statistical inference should provide tools for the understanding of typical functional connectivity patterns, as well as quantification of familiar concepts like sample and population variability, and dependence on clinical phenotypes via regression. Even though some progress in the direction of population level inference has recently been made in the context of fMRI data (\citealt{Narayan:2015}; \citealt{Shou:2014}), typical analyses are still reliant on untenable assumptions of time-independence. The literature is, in fact, substantially silent on the subject of population level connectivity inference  using EEG data. In this work, we aim to address this problem and introduce a simple and interpretable technique for the analysis of brain synchronicity from a sample of EEG readings. Our approach relaxes the classical graphical modeling strategy into a simpler problem of clustering brain regions. As a consequence our analysis is perhaps coarser than what is usually done in the functional connectivity literature.

Our approach is based on the definition of cortical maps, identifying areas of synchronous neuronal activity specific to individual subjects and experimental epochs, intended as time intervals. Synchronized cortical regions are estimated via a mixture model of eigen-Laplacian vectors, obtained from appropriately constructed dissimilarity matrices.   As the experiment evolves in time, subject and time-specific cerebral maps form a longitudinal ensemble. In this context, we posit that pooled information, within and between subjects, is amenable to statistical analysis via a hierarchical model involving mixture probabilities (\citealt{Lock:2013}), which we call Multilevel Integrative Clustering (MIC).  Our framework supports both the definition of coordinated neuronal activity via a mixture approach, and the formulation of probability statements describing inter-subject and intra-subject variability via the familiar toolset of hierarchical modeling.

Our manuscript is organized as follows. In Section \ref{sec:MIC} we describe a general framework for integrative clustering at the epoch, subject and population levels. In Section \ref{sec:Simulation} we assess the operative characteristics of our proposed approach through experiments on engineered data. In Section \ref{sec:Case Study} we apply the proposed framework to the analysis of a resting-state EEG study on typically developing (TD) children and children diagnosed with Autism Spectrum Disorder (ASD). We conclude with a critical discussion and potential extensions in Section \ref{sec:Discussion}.

%% METHODS --------------------------------------------------------------------------------------

\section{Multilevel Integrative Clustering (MIC)}
\label{sec:MIC}

In the following discussion we proceed to characterize coordinated neuronal activity via time-varying pairwise distances between the time series associated with a set of EEG sensors or electrodes. Our approach builds on  \cite{euan2015spectral}, who define synchronicity in relation to pairwise similarities between the power spectral densities of electrode-level signals.  In \S \ref{subsec: preprocessing}, we describe a data meta-processing step aimed at obtaining stable time-varying estimates of the EEG spectral profiles. In \S \ref{sec: model}, instead of directly operating on spectral densities, we model a set of related $d$-dimensional eigen-Laplacians via a multilevel model for clustering areas of synchronous neuronal activation.  Inferential and computational details are discussed in \S \ref{subsec:postinference} and \S \ref{sec: number cluster}.

\subsection{From EEG Signals to eigen-Laplacian Matrices}
\label{subsec: preprocessing}
The spectral analysis of neural signals is an important workhorse in EEG studies, as frequency bands are thought to be associated with specific cognitive, perceptive and cellular phenomena (\citealt{Teplan:2002}). EEG time-series signals are usually collected in relation to a geodesic net of $p$ electrodes.   Upon collection, raw signals are segmented into 1024ms  time intervals for EEG preprocessing,  which typically includes bandpass filtering, electrode and segments rejection, and artifacts inspection. Similar pipelines are common for EEG analysis, which can improve the SNR for spectral analysis (\citealt{Bigdely-Shamlo:2015aa}).

We are interested in the time-dynamics of neuronal synchronicity through a notion of time-varying spectral density via local stationarity. 
More precisely, we consider a sequence of stochastic processes $\{Z_{t, T},\; t=1,2,\ldots, T, \;T\in \mathbb{N}\}$ to be locally stationary in the sense of \cite{Dahlhaus:1997}-definition 2.1. Assuming a smooth transfer function characterizing the stochastic evolution of $Z_{t, T}$, the time-varying spectral density of the process is defined as:
$$\phi_Z(\omega, t) := \frac{1}{2\pi}\sum_{\tau=-\infty}^\infty \mbox{cov}\left\{Z_{t, T},\,Z_{(t+\tau, T)}\right\}\mbox{exp}(-i\tau\omega), \;\; \omega\in [0,\pi].$$
Intuitively, $\phi_Z(\omega, t)$ may be interpreted as the variance contributed to the series, in a neighborhood of $t$, by oscillations in a narrow frequency band around $\omega \in [0,\pi]$.

Let $i=1,\ldots, n$ index $n$ study subjects, $j=1,\ldots, p$ index $p$ EEG electrodes, and $s_{\ell i}$, $\ell_i = 1,\ldots, q_i$, index $q_i$-1024ms segments retained after data quality control.  The filtered EEG data can be seen as an ensemble of time-series segments $Y_{ij}(s_{\ell_i})$, each composed of a number of measurements reflective of analog-to-digital sampling rates, typically 256/512Hz.

In our formulation we fully acknowledge common pre-processing practices, which sees qualifying EEG segments  being concatenated and re-referenced without time labelling. This practice typically leads to latent gaps in the post-processed series, providing a non-standard inferential framework for time-varying spectral estimation.

In order to obtain time-varying stable estimates of electrode-specific spectra, we operate on a combined set of $\gamma$ adjacent segments  $(s_{\ell_i}, \ldots, s_{(\ell_i + \gamma)})$, which we define as epochs.  Furthermore, adjacent epochs smooth over the original time domain by overlapping over a $\delta \in (0,1)$ fraction of segments. For each subject $i$, electrode $j$ and epoch $t \in \{1,2,\ldots,T_i\}$, we obtain estimates $\widehat{\phi}_{ij}(\omega, t)$ of the epoch-specific spectral density by averaging segment specific spectral density estimates obtained as in \cite{Ombao:2001}. The details of this procedure are reported in a supplementary document. Our approach stems from the idea introduced by \cite{Hasenstab:2015} in the context of time-domain analyses, and approximates the continuous time spectral analysis reported in \cite{rosen2012adaptspec}. The use of overlapped sliding windows in the estimation of a time-dependent power spectral density mediates between the need for stable estimates and the potential for non-stationarity over the entire duration of the study.  A study of inferential robustness to smoothing choices is reported in \S \ref{sec:Simulation}.

Following the approach by \cite{euan2015spectral}, desynchronicity is measured by {\em total variation distance} (TVD) between a pair of normalized  spectral densities estimated at each epoch, so that, for subject $i$, desynchronicity between electrode $j$ and electrode $k$ at epoch $t$ is  defined as: 
\[ d_{it}(\widehat{\phi}_{ij}, \widehat{\phi}_{ik}) = 1-\int \min \{\widehat{\phi}_{ij}(\omega, t),\,\widehat{\phi}_{ik}(\omega, t)\} \dd \omega.\] 
For each subject and epoch, these pairwise distances produce a $p\times p$ dissimilarity matrix 
$D_i(t) = \left[d_{it}(\widehat{\phi}_{ij}, \widehat{\phi}_{ik})\right]$, 
summarizing information on differential synchronicity between the $p$ electrodes from different cortical regions.  

Before clustering, each matrix is represented in the eigen-space spanned by the largest $d$ eigenvectors of the graph-Laplacian associated with an affinity matrix $A_i(t) = 1-D_i(t)$. More precisely, we take a {\it graph cuts} view of clustering and construct a normalized graph-Laplacian $\mathcal{G}_i(t) = diag\left[A_i(t){\bf 1}_p\right]^{1/2}\,A_i(t)\,diag\left[A_i(t){\bf 1}_p\right]^{1/2}$, representing  a weighted undirected graph between EEG electrodes.  In this setting, we follow (\citealt{Ng:2001}) and summarize the information in $\mathcal{G}_i(t)$ with its largest $d$ eigenvectors $X_i(t) \in \mathbb{R}^{p\times d}$. 

This strategy is intuitively motivated by the analysis of the  isolated connected components ``{\it ideal case}'', in which $A_{jk}(t) > 0$ iff components $j$ and $k$ belong to the same cluster, and $A_{jk}(t) = 0$ otherwise. In this simplified setting, considering $K$ clusters, the first $K$ columns of $X_i(t)$ have non-zero elements corresponding to connected components in $A_i(t)$. Row-wise, $X_{i}(t)$ is piece-wise constant, suggesting K-means as a simple clustering rule to recover the connected components. 

We work under the assumption that $\mathcal{G}_i(t)$ is a perturbation of the ``{\it ideal case}'' and in \S \ref{sec: model} we exploit this intuition to develop model-based clustering of electrodes at the epoch, subject and population level.  
Crucially, we avoid using a mixture model of  spectral densities; instead model-based clustering of EEG signals over potentially non-convex manifolds is achieved using simpler location/scale-mixture  models involving vectors in $\mathbb{R}^d$. 

It is important to point out, that the measure of neuronal synchronicity, defined as spectral synchronicity, is indeed not essential for the application of multilevel integrative clustering. In particular, alternative means of quantifying similarity between time series, like coherence, cross correlation, partial correlation, etc. may be appropriate in specific investigations (\citealt{Bastos:2016}). Furthermore, if interest centers on specific band-power frequencies, discrepancies are easily defined over the appropriately truncated spectral densities.

\subsection{Hierarchical Mixture Priors and Multilevel Inference}
\label{sec: model}
Let $X_{ij}(t) \in \mathbb{R}^d$, be a $d$-dimensional eigen-Laplacian vector associated with the EEG signal for subject $i$, ($i=1,2,\ldots,n$); electrode $j$, $(j=1,2,\ldots,p)$; at epoch $t = 1,2,\ldots,T_i$. 
In practice, we observe subject-specific epochs $t_{im_i}$, $(m_i=1,2,\ldots,T_i)$. However, without loss of generality and for ease of notation, we maintain the lighter epoch indexing $t$ throughout the manuscript.

Within subject, at epoch $t$, we conceptualize synchronous patterns of cortical activity, by clustering electrodes according to the following mixture model.
Denoting with \linebreak $f\{\cdot \mid \cdot\}$ a generic density with respect to the Lebesgue measure on $\mathcal{B}\left(\mathbb{R}^d\right)$, we assume that 
each eigen-Laplacian vector $X_{ij}(t)$ is sampled from a $K$-components mixture distribution, indexed by parameters $\bth_{ik}(t)$ and mixture probabilities $p_{ijk}(t) \in [0,1]$, such that: 
\begin{equation}
\label{eq: sampling model}
X_{ij}(t) \sim \sum_{k=1}^K p_{ijk}(t) f\{X_{ij}(t)\mid \bth_{ik}(t)\},  \hspace{0.5cm} \sum_{k=1}^K p_{ijk}(t) = 1.
\end{equation}
We find it convenient to re-express this sampling model with the equivalent hierarchical representation, mixing over cluster labels  $L_{ij}(t) \in \{1,2,\ldots,K\}$, s.t.:
\begin{equation}
\label{eq: sampling model hierarch}
\begin{array}{ccl}
X_{ij}(t)\mid L_{ij}(t) = k &\sim& f\{X_{ij}(t)\mid \bth_{ik}(t)\},\\
\Pr\{L_{ij}(t)= k\} &=& p_{ijk}(t).
\end{array}
\end{equation}
In this setting, echoing the clustering ``{\it ideal case}'' discussed in the previous section,  we exploit the connection between $K$-means and Gaussian mixtures and represent the sampling density in (\ref{eq: sampling model}) as a $K$-component  location/scale mixture of Gaussian distributions. Specifically, let $\mu_{ik}(t) \in \mathbb{R}^d$ be a $d$-dimensional mean vector, and $\sigma^2_{ik}(t) > 0$ be a variance parameter. We assume: 
\begin{equation}
\label{eq: likelihood}
f\{X_{ij}(t)\mid \bth_{ik}(t)\} = N\{\bmu_{ik}(t), \sigma^2_{ik}(t)I_d\}.
\end{equation}
Given the sampling model in (\ref{eq: sampling model hierarch}), our proposed approach for the integration of
information at the subject and population levels follows a conceptually simple  strategy, building directly on the setting of multilevel modeling (\citealt{gelman:2007}). Crucially, we maintain that mixture means and variances are independent across subjects and epochs, but posit that cluster configurations,  conceptualizing synchronicity of brain regions, are likely to adhere to patterns of similarity within and between subjects. 

We make this idea precise by specifying a hierarchical prior for the mixture probabilities, $p_{ijk}(t)$. This is achieved by defining conditionally exchangeable mixture configurations, where epoch-level clusters $L_{i}(t)$ are obtained, {\it a priori}, as a stochastic perturbation of a time stable subject-level clustering, indexed by $C_i$. Similarly, subject level configurations, $C_i$, are obtained as a stochastic perturbation of a population-level cluster, indexed by $S$.

Let $C_{ij} \in \{1,2,\ldots,K\}$ be  the cluster label for electrode $j$ at the level of subject $i$. Furthermore, let $\beta_i(t) \in [1/K, 1]$ be an adherence parameter, quantifying conformity between cluster assignments at epoch $t$ and the subject-level label $C_i = (C_{i1},\ldots, C_{ip})'$. We assume, 
\begin{equation}
\label{eq: subject clust}
\Pr\{L_{ij}(t )= k \mid c_{ij}\}  \,\equiv\, \nu_c\{k,c_{ij},\beta_{i}(t)\} \,=\, \left\{\begin{array}{ll}
\beta_i(t) & \mbox{ if } c_{ij}=k\\
\frac{1-\beta_i(t)}{K-1} & \mbox{ otherwise}\\
\end{array}\right. ,
\end{equation}
where the probability $\nu_c\{\cdot,\cdot,\cdot\}$ is defined implicitly. This prior defines a probabilistic anchor, relating epoch level patterns of synchronicity at the subject level via simple and interpretable parameters $\beta_{i}(t)$. The underlying assumption is that epoch-level patterns of synchronicity are allowed to vary dynamically with $t$, but that variation in cluster configurations is anchored at the subject-level by a consensus pattern $C_i$. 

A similar anchoring strategy is pursued at the population level. Specifically, let $S_j\in \{1,2,\ldots,K\}$ be a population level cluster label for electrode $j$, and $\alpha_i\in [1/K, 1]$ be an adherence parameter, quantifying conformity between cluster assignments for subject $i$ and population level labels $S = (S_1,\ldots, S_p)'$. We assume, 
\begin{equation}
\label{eq: pop clust}
\Pr(C_{ij} = k \mid s_j)\, \equiv\, \nu_s(k,s_j,\alpha_{i}) \,=\, \left\{\begin{array}{ll}
\alpha_i & \mbox{ if } s_j = k\\
\frac{1-\alpha_i}{K-1} & \mbox{ otherwise}\\
\end{array}\right. ,
\end{equation}
where probability $\nu_s(\cdot,\cdot,\cdot)$ is defined implicitly. The model is completed by specifying population level prior proportions:
$$\Pr(S_j = k) = \pi_k, \;\;(k=1,2,\ldots, K).$$
To build intuition about the nature of these priors, we note that, if $\alpha_i=1$, we expect cluster assignments for subject $i$ to match exactly the population-level labels with probability 1. In contrast, for $\alpha_i$ approaching the value $1/K$, electrode clustering configurations $C_i$, for subject $i$, are drawn independently of the population level labels $S$. Similar considerations apply to  $\beta_i(t)$, as these paratmeters relate subject- and epoch-level cluster configurations.

This modeling strategy is loosely related to the idea of consensus clustering (\citealt{Nguen:2007}), as applied to the integration of multi-source data. Our specific formulation is a direct generalization to multilevel models of the approach taken by \cite{Lock:2013} to the integration of heterogeneous genomic data.

In our multilevel setting, the conditional posterior distribution for epoch-level cluster labels $L_{ij}(t)$ is easily defined as: 
\begin{equation}
\label{eq: post epoch}
\begin{array}{lcl}
\Pr\{L_{ij}(t) = k \mid X_{ij}(t), c_{ij}, s_j, \bth_{i}(t)\} & \propto & f\{X_{ij}(t)\mid L_{ij}(t) = k,\bth_{i}(t)\}\Pr\{L_{ij}(t) = k \mid c_{ij}\}\\
&= & f\{X_{ij}(t) \mid \bth_{ik}(t)\}\nu_c\{k,c_{ij},\beta_i(t)\}.
\end{array}
\end{equation}
This form highlights how inference on $L_{ij}(t)$ integrates information from both data $X_{i}(t)$ at epoch $t$, and subject-level clustering $C_i$ (assumed stable across epochs), through a weighting scheme proportional to the size of the adherence parameter $\beta_{i}(t)$.

At the subject-level,  conditional posterior probabilities of cluster membership weigh  epoch level configurations $L_{i}(t) = (L_{i1}(t),L_{i2}(t),\ldots,L_{ip}(t))'$ with population level configurations $S$, through adherences $\alpha_i$ as follows: 
\begin{equation}
\label{eq: post subject}
\begin{array}{lcl}
\Pr\{C_{ij} = k \mid \ell_{ij}(1),\ldots,\ell_{ij}(T_i), s_j\} &\propto & \Pr\{\ell_{ij}(1),\ldots,\ell_{ij}(T_i) \mid C_{ij} = k\} \Pr\{C_{ij} = k \mid s_j\}\\
&= & \prod_{t=1}^{T_i} \nu_c\{\ell_{ij}(t),k,\bbeta_i\}\,\nu_s(k,c_j,\alpha_i).
\end{array}
\end{equation}
Finally, at the population level, overall consensus labels $S$ are determined according to the following conditional posterior probability:
\begin{equation}
\label{eq: post population}
\Pr(S_j = k \mid c_{1j},\ldots,c_{nj},\Pi,\balpha) \propto \pi_k\prod_{i=1}^n \nu_s(k,c_{ij},\alpha_i).
\end{equation}
In summary, for each subject we infer a consensus cortical configuration $C_i$, combining epochs $L_i(t)$ trough coherence weights $\beta_i(t)$. Across subjects, group-level inference is conducted through a consensus configuration $S$, pooling subject-level configurations $C_i$ through coherence parameters $\alpha_i$.  This stochastic structure allows for a very general conceptualization of dependence across epochs. We note that, in some cases, more structured priors could be warranted, including a fully exchangeable configuration, where $\beta_i(t) = \beta_i$, for all epochs; or the case of $\beta_i(t)$ being defined as a smooth function of the epoch index $t$.  Similar restrictions may be applied to $\alpha_i$, depending on inferential goals and experimental characteristics.

\subsection{Posterior Inference}
\label{subsec:postinference}
We discuss posterior inference for the model in $\S$ \ref{sec: model} on the basis of MCMC samples from the target distribution. Even though multilevel modeling of cluster labels is a somewhat non-standard approach in a hierarchical setting, conditionally conjugate analysis is indeed possible, resulting in  significant simplifications in computation and inference. 

Specifically, we consider a standard Dirichlet prior for population-level proportions, so that $\Pi = (\pi_1, \pi_2, \ldots, \pi_K)' \sim \mbox{Dirichlet}(\eta)$. Epoch-level means and variances, are chosen to be conjugate to the
graph Laplacian likelihood in (\ref{eq: likelihood}). Letting $\bth_{ik}(t_{im}) = (\bmu_{ik}(t_{im})', \sigma^2_k(t_{im}))'$, we assume that $\bth_{ik}(t_{im}) \sim N\Gamma^{-1}(\bmu_0, \lambda_0, \xi_{01}, \xi_{02})$. Finally, 
subject-level adherence parameters $\alpha_i$ and epoch-level adherence parameters $\beta_i(t_{im})$ are assigned
truncated Beta priors, with left truncation at $1/K$, so that:
$$\alpha_i \sim \mbox{TBeta}(a_i,b_i,1/K),  \mbox{ and } \;\; \beta_i(t) \sim\mbox{TBeta}(c_i,d_i,1/K).$$
A justification for these truncated Beta priors may be obtained by considering the form of the marginal allocation probabilities at subject and epoch level.  Given $\Pi$, subject-level allocation probabilities are expressed as:
\[
p_{ik} = \Pr(C_{ij}=k\mid \pi_k) = \pi_k \alpha_i+(1-\pi_k)\frac{1-\alpha_i}{K-1}.
\]
Similarly, at the epoch level, we have:
$$
\Pr\{L_{ij}(t) = k \mid \Pi\}= \sum_{c_{ij}} \Pr\{L_{ij}(t) = k \mid c_{ij}\} \Pr(c_{ij}\mid\Pi) 
                        = \beta_i(t) p_{ik} + (1-p_{ik})\frac{1-\beta_i(t)}{K-1}.
$$
At both levels, an adherence value of $1/K$ corresponds to allocation probabilities, which are independent of higher-level clustering realizations. 

A Gibbs sampler targeting the posterior distribution is easily devised, by iterating through a transition sequence of full conditional posteriors. Specific details about the form of the conditional posterior densities are reported in a supplementary document.

At each level of the model, the posterior probability associated with set of clustering labels, for generality say $p(C \mid {\bf Y})$, and the corresponding MCMC samples, summarize our knowledge about potential partitions of cortical regions into synchronously activated areas. Based on the information in this posterior, we may be interested in selecting a representative partition, say $C^*$. Following \cite{dahl2006model}, we avoid using the na\"{i}ve maximum {\it a posteriori} (MAP) estimate and instead consider a point estimator based on least squares. More precisely,  consider an MCMC sample of $M$ $p-$dimensional label configurations, $\{C^{(r)}: \; r=1,2,\ldots,M\}$. For each sample, we define a $p\times p$ adjacency matrix $\mathcal{A}\left(C^{(r)}\right) = \left[\mathcal{A}\left(C^{(r)}\right)_{ij}\right] =\left[I(C^{(r)}_i = C^{(r)}_j)\right]$. Let $\bar{\mathcal{A}}$ be an estimate of the posterior mean  $E[A \mid Y]$.
The least square estimate $C^*$ is selected from posterior realizations which minimize the following Frobenius norm
\[  C^* = \min_{C^{(r)}, r=1,\ldots,M} ||\, \mathcal{A}\left(C^{(r)}\right) - \bar{\mathcal{A}} \,||_2.\]
Uncertainty about clustering estimates can be obtained from the posterior distribution, locally by quantifying pairwise relative frequencies of  synchronization or globally via the distribution of $\mathcal{D} = || A^{(r)} - \bar{A}||_2$. 
Examining this quantity facilitates direct comparison between subject and population level clustering results, allowing for low dimensional assessment of cluster quality, population and individual-level variability. 

Computation and inference for MIC is performed under the R environment. A readily compiled package is available from the corresponding author's GitHub page.

\subsection{Number of Clusters and Identifiability}
\label{sec: number cluster}
Posterior inference as described in \S \ref{subsec:postinference} presumes a known number of clusters $K$ and a known number of eigen-Laplacian components $d$. For given $d$, selection of the number of mixture components, $K$, may be based on  information criteria. In our simulation studies we find that the Bayesian Information Criterion (BIC) (\citealt{schwarz1978estimating}) tends to outperform more complicated indices. Our findings are in agreement with   \cite{steele2010performance}, who observed that BIC outperforms  many other criteria including ICL, DIC, and AIC, especially in the case of Gaussian mixture models.

The choice of $d$ is less trivial, even though, some theoretical results point to the inclusion of the first $K$ eigenvectors as being sufficient in the task of separating $K$ groups, (\citealt{Ng:2001}). Guided by this general principle, we perform a joint search on the dimensionality of the eigen-Laplacian $d$, and the number of clusters $K$ simultaneously. More precisely, within a specific dimension $d$, the optimal value of $(K \mid d)$ is determined by the maximal BIC. Starting from low dimensions, usually $K=d=2$, we allow for up-transitions on dimensionality, when $K^*\mid d > d$. Stopping rules, aiming at achieving stable solutions around the equality of $d^{*} = K^{*}$ are determined heuristically. Details  are reported in Algorithm~\ref{alg:dkselect}. Crucially, we avoid complete enumeration over all $(d, K)$ combinations, and propose a search strategy which is linear in the maximum number of clusters. Our empirical studies in \S \ref{sec:Simulation} show good performance  and fast convergence to well behaved solutions.

%% Algorithm
\begin{algorithm}
\caption{$(d,K)$ Selection} \label{alg:dkselect}
\begin{algorithmic}[1]
	\State Set $d = 2$, $K = 2$;
	\State $\text{current\_BIC} = \text{BIC}(d,K)$;
	\While{$d\le $max\_$d$}
		\While{BIC$(d,K+1) \ge$ current\_BIC}
			\State current\_BIC = BIC$(d,K+1)$;
			\State $K = K+1$;
		\EndWhile
		\If{$d \ge K$}
			\State {\bf break};
		\Else
			\State $d = K$,  $K = K-1$;
			\State $\text{current\_BIC} = \text{BIC}(d,K)$;
		\EndIf
	\EndWhile
	\State \Return{$(d,K)$}
\end{algorithmic}
\end{algorithm}

\vskip.1in

\noindent For given $d$ and $K$, simulation based procedures, including MCMC, are usually prone to {\it label switching} (\citealt{celeux2000computational}). In the setting of the model proposed in \S \ref{sec: model} the same phenomenon may occur both within and between data levels. An important aspect of simulation-based inference in multilevel clustering is, therefore, the enforcement of correspondence between component labels of epochs, subjects and population level clustering. Possible remedies include artificial identifiability constraints, relabeling procedures, and label invariant loss functions (\citealt{jasra2005markov}). Within the multilevel setting, we proceed with online class relabeling or alignment. More precisely, we operate within population and subject-level indexes to find permutations of labels that maximize  adherence with the population level clustering. Specifically, all newly sampled labels are permuted to insure maximal alignment with the population indexes. If $\mathcal{A}_0$ is an adjacency matrix  as defined in \S \ref{subsec:postinference}, representing the current state of the population level labels $S$, and $\mathcal{A}_q$ is an adjacency matrix representing the current state of any other level clustering, optimal alignments are obtained by maximizing $tr\left(\mathcal{A}_0^\prime \mathcal{A}_q\right)$ over $k!$ possible permutations.

%
% Simulations -------------------------------------------------------------------------------------------
%
\section{ Monte Carlo Studies}
\label{sec:Simulation}
To investigate the operating characteristics of the proposed framework, we simulate EEG signals with the desired oscillation features from a mixture of AR(2) processes. We seek to evaluate: (1) the sensitivity of MIC results to differing sliding window size, $\gamma$, (2) the accuracy of estimated quantities for varying group adherence, (3) the performance of the model selection strategy proposed in Algorithm~\ref{alg:dkselect},  and (4) the behavior of population level clusters under varying signal to noise ratio (SNR) and varying sample size.

\subsection{Simulation setup for spectrally specified EEGs}
\label{subsec:EEG simulation}

We make an effort to tailor the simulation of engineered time series in a way that mimics a sample of EEG readings typically seen in practice. To this end, we note that EEGs are often expected to feature oscillation patterns at different frequency bands: delta (0.5-4 Hz), theta (4-8 Hz), alpha (8-12 Hz), beta (12-30 Hz) and gamma (30-50 Hz). Waveforms that are subdivided into bandwidths are thought  to correspond to region-related activities on the cortex, both normally and pathologically. 

\begin{figure}
\begin{center}
\begin{tabular}{ccc}
\small  (a)  \footnotesize Main-state & \small (b) \footnotesize Off-state & \small (c) \footnotesize Spectral realization\\
\includegraphics[width = 0.315\linewidth]{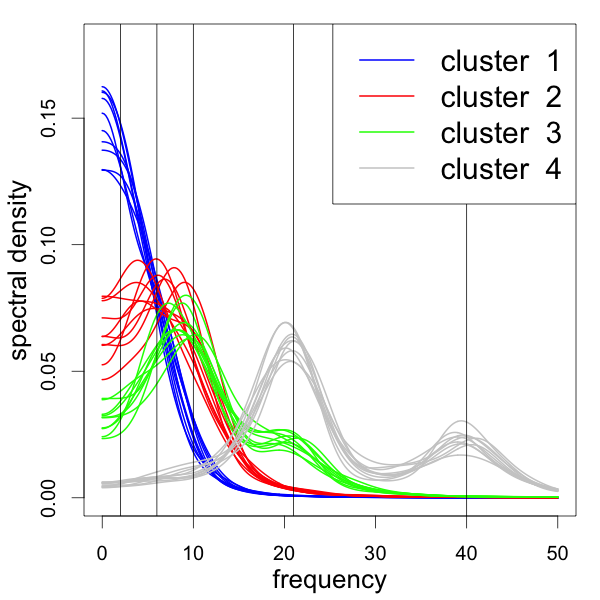}&
\includegraphics[width = 0.315\linewidth]{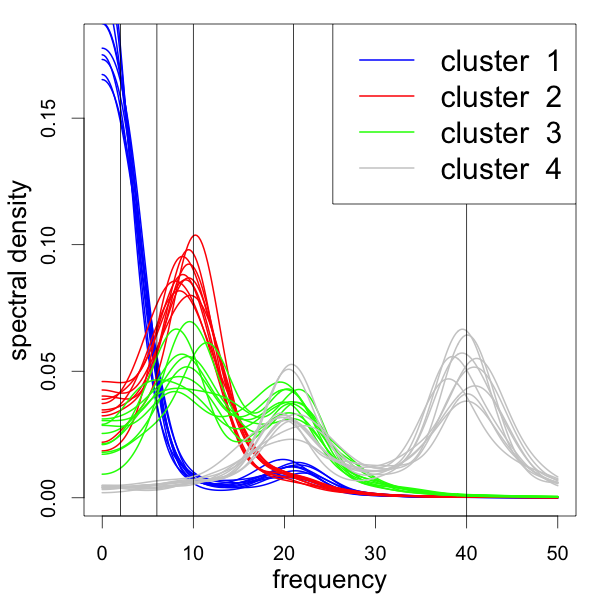}&
\includegraphics[width = 0.315\linewidth]{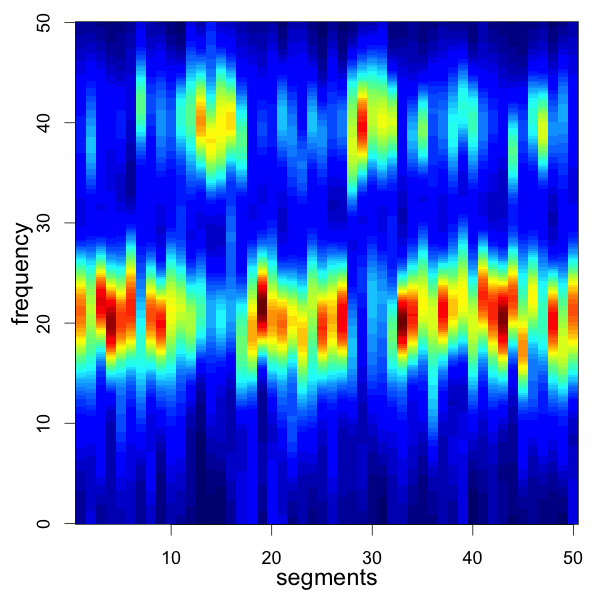}\\
\end{tabular}
\end{center}
\caption[Simulated Spectra]{\small {\bf Simulated spectral configurations}:  (a) {\it main-state}  spectral densities. (b) {\it off-state} spectral densities. (c) Segment-by-segment normalized power spectral densities for a piecewise stationary process simulated from cluster 4.}
\label{fig:spec}
\end{figure}

Our strategy, aims to simulate this spectral distinguishability by allowing each  spectrum to exhibit concentrated (peak-shaped) energy in at most two frequency bands. Given a family of spectra, EEG time-series are simulated from a linear mixture of second order auto-regressive AR(2) processes. Details about the data generating mechanism are reported in a supplementary document. Furthermore, we represent potential non-stationarity by generating time-series as realizations from a piecewise stationary process, alternating randomly between two spectral configurations:  a \emph{main}-state (Fig \ref{fig:spec}(a)), and an  \emph{off}-state shown in Fig \ref{fig:spec}(b).  The \emph{main}-state has a time span $t_{\text{main},i}\sim$ exp$(\lambda)$, with $\lambda = .05$s, followed by the \emph{off}-state which has a time span $t_{\text{off},i}\sim N(5,1)$. Fig \ref{fig:spec} (c) depicts  this piecewise-stationarity for one electrode from the simulated samples. Cluster labels are generated as follows:

\begin{enumerate}
\item At the {\it population level}, we structure cluster labels $S_j, \;\;(j=1,\ldots,p=100)$ to partition $100$  sensors into 4 balanced clusters.
\item Draw $\alpha$ from a Uniform$(0.5,1)$ distribution. For each subject $i$, and $j = 1,\ldots, p=100$; generate {\it subject level} labels $C_{ij} \in \{1,2,3,4\}$ with probabilities $\Pr(C_{ij} = S_j) = \alpha$ and $\Pr(C_{ij} \ne S_j) = (1-\alpha)/3$.
\item Given $C_{ij}$, generate piecewise stationary processes for 50 seconds, according to the {\it main}-state / {\it off}-state mechanism described previously. 
\end{enumerate}
\noindent Our Monte Carlo study is based on 100 datasets. Subject-level variation is induced semi-parametrically, via random reconfigurations of subject specific clusters, and random timing of the {\it main/off}-state segments.   The number of subjects, electrodes and segments were chosen to mimic the sampling structure in our case study. Note that in this setting, knowledge of the timing of {\it main}-state, {\it off}-state would result in perfect agreement of cluster labels within subject. Our simulation is therefore engineered to detect specific sensitivity to alternative metapreprocessing strategies.

\subsection{Operating characteristics}
\label{subsec:simout}
\begin{figure}
\centering
\begin{tabular}{ll}
\small (a) & \small (b) \\
\includegraphics[width = 0.46\linewidth]{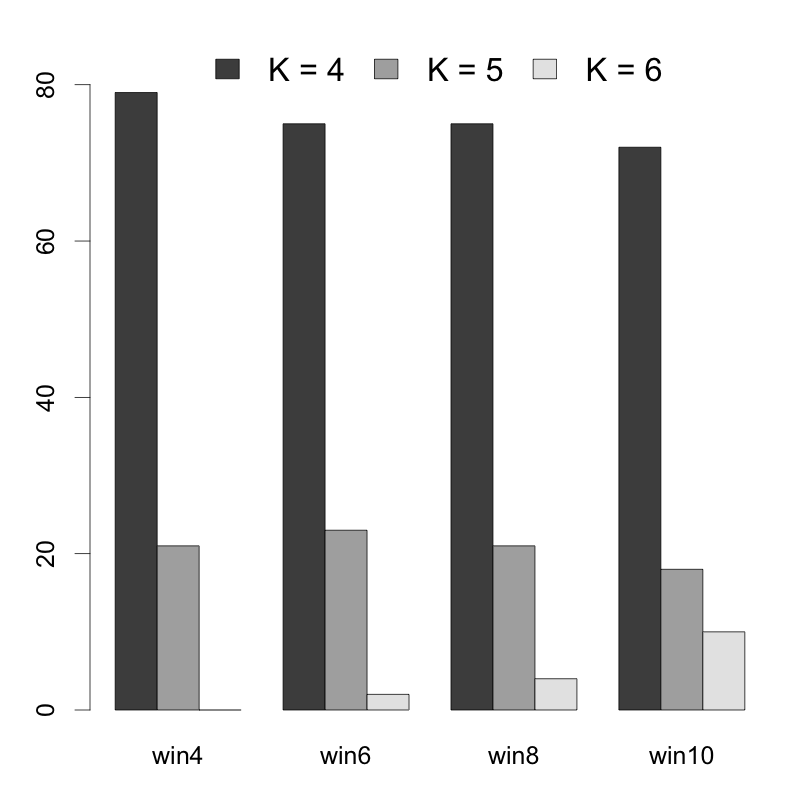}&
\includegraphics[width = 0.46\linewidth]{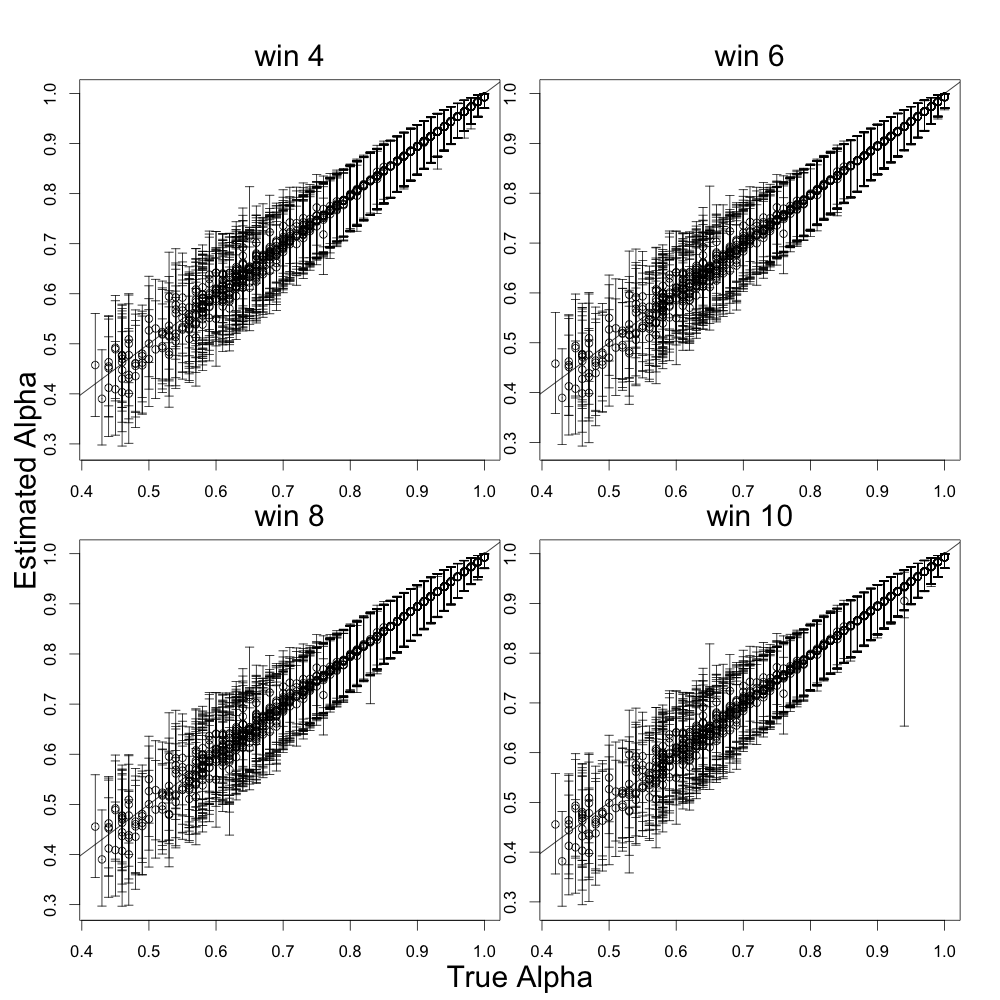}\\
\small (c) & \small (d)\\
\includegraphics[width = 0.46\linewidth]{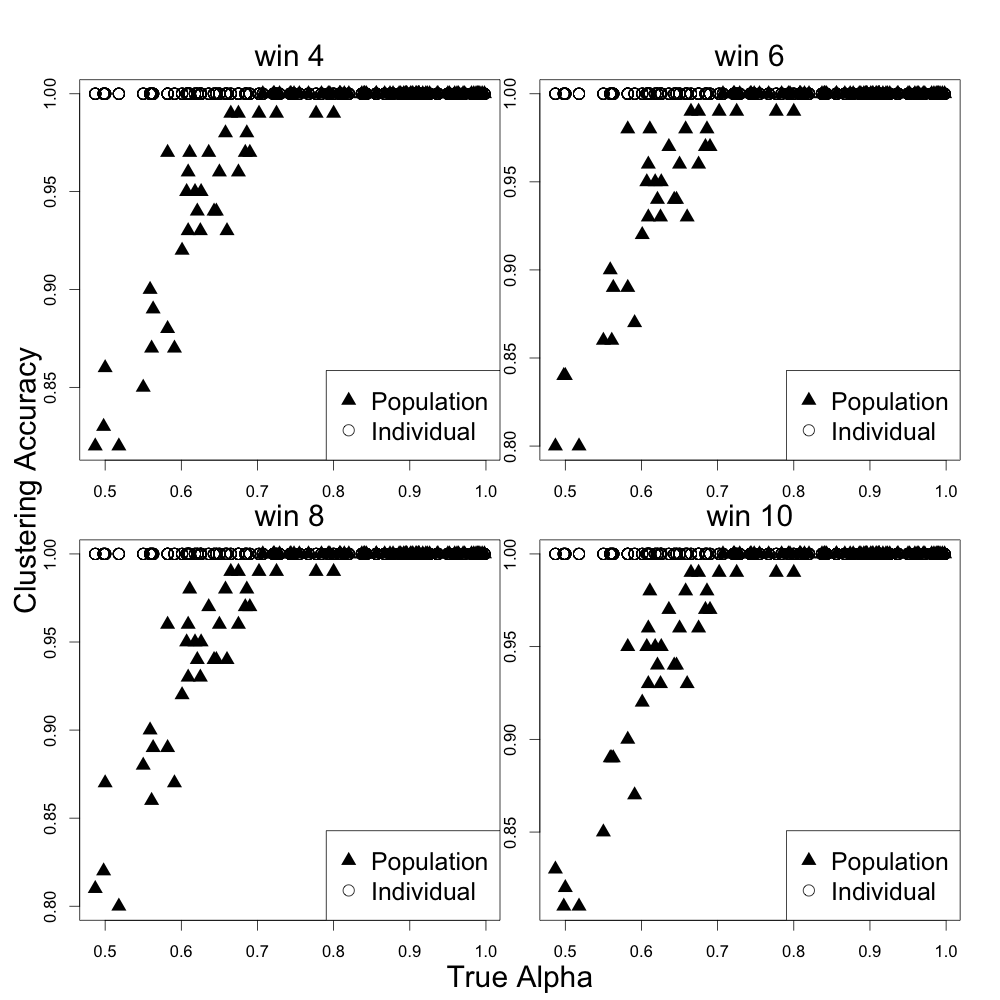}&
\includegraphics[width = 0.46\linewidth]{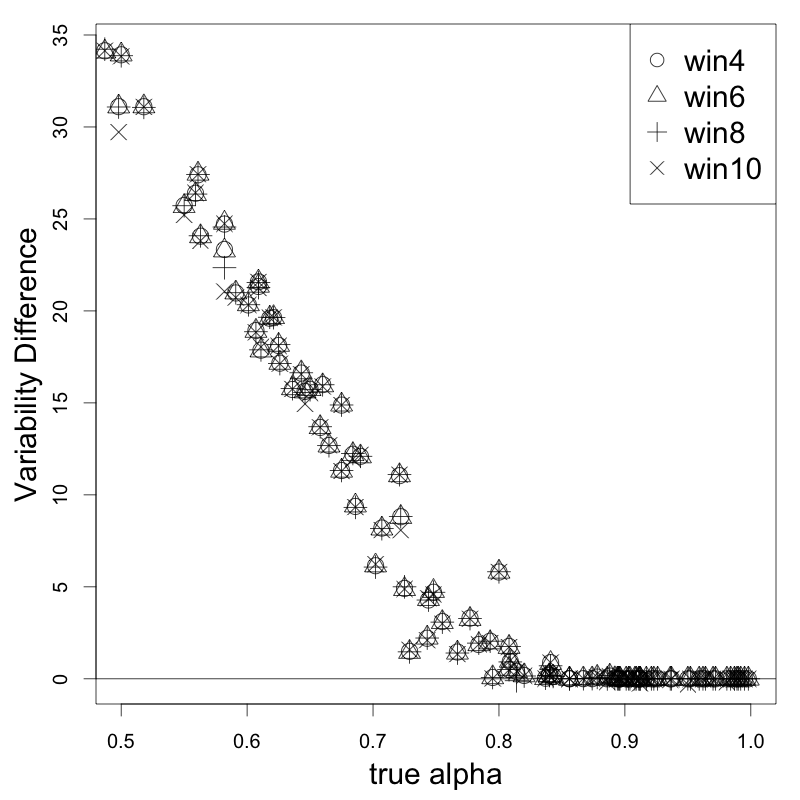}\\
\end{tabular}
\caption[Simulation Results]{\small {\bf Simulation results:} 
(a) Path-length for the search in Algorithm 1 for varying smoothing configurations in $\gamma$.
(b) Estimated adherence parameters $\widehat{\alpha}$'s and 95\% credible intervals against the data generating truth.
(c) Clustering accuracy against generating $\alpha$'s at the subject- and population-level.
(d) Average difference in clustering variance against true $\alpha$'s.}
\label{fig:simout}
\end{figure}

In \S \ref{subsec: preprocessing} we introduced a pre-processing step to smooth over the duration of the EEG recordings in order to obtain time-stable estimates of spectral densities.
We start by assessing sensitivity of window size, $\gamma \in \{4,6,8,10\}$, at a fixed $\delta=0.5$ fraction of overlap between epochs. Algorithm~\ref{alg:dkselect} successfully selected the correct number of clusters ($K=4$), in more than 76\% of cases for all varying widow sizes, Fig~\ref{fig:simout}(a).

Furthermore, we investigate the performance of MIC  under varying degrees of subject-specific variability, by examining estimates of adherence between subject- and population-level clustering.   Fig~\ref{fig:simout}(b) depicts posterior medians $\widehat{\alpha}_i = E(\alpha_i \mid {\bf X})$ and their 90\% credible intervals, based on the 5 and 95 percentiles, against the true $\alpha$'s. Posterior estimates are generally close to their true values, and over 99\% of the credible intervals cover the true $\alpha$'s. 

Clustering accuracy, defined as the percentage of correctly classified electrodes, is assessed both at the subject and population level, Fig~\ref{fig:simout}(c). Estimated subject-level clusters tend to be recovered accurately ( $>$ 98\%), regardless of $\alpha$ values. As expected, accuracy in the recovery of population level patterns relies on the magnitude of subject-level adherence to the population, with accuracy approaching 100\% as $\alpha \rightarrow 1$. 

Finally, we investigate the relationship between subject-level and population-level clustering variance estimates as a function of adherence and meta-processing strategy, Fig~\ref{fig:simout}(d). Our summaries focus on a measure of global variance $\mathcal{D}$, as defined in \S \ref{subsec:postinference}. More precisely, denoting the clustering variance by $\mathcal{D}_S$ at the population level, and by $\mathcal{D}_{C_i}$ at the level of subject $i$, we consider the average difference in clustering variance, defined as: 
$
\Delta_{\mathcal{D}} = E(\mathcal{D}_S\mid {\bf X}) - \frac{1}{n}\sum_i E(\mathcal{D}_{C_i} \mid {\bf X}).
$
As the adherence simulation truth approaches a level of complete agreement ($\alpha\rightarrow 1$), the average difference in clustering variance $\Delta_{\mathcal{D}}$ converges to zero, indicating that average subject-level and population-level cluster variances reach similar magnitudes over strongly adherent clustering patterns, Fig~\ref{fig:simout}(d). 

A second set of simulation studies aims to assess the operative characteristics of the proposed method under different SNR and sample size settings. Specifically, we consider SNR = $1,5,10$ and sample size $N=10,20,40$. We assess performance of our method under stationarity and local stationarity. Details about the simulation procedure are reported in our supplementary materials document. Our experiments show that group-level inference is highly robust to SNR configurations. In both stationary and locally stationary settings, clustering accuracy increases with sample size, going from a minimum of 0.8 (N=10), to about 0.9 (N=40). Results are reported in Table~\ref{tab:simulation}. 

From our experiments we conclude that estimation and clustering results tend to be robust across a broad range of SNR and smoothing parameters. This feature is likely to be useful in many applications, where it is usually hard to develop meta-processing gold standards.

%
%% Case study and Data Analysis ------------------------------------------------------------------------------
%
\section{A Case Study on Resting State Brain Activity}
\label{sec:Case Study}
Our study originates from an experiment aimed at understanding children's neurocognitive development. The study was carried out in the department of Psychiatry at UCLA and aims to cluster spectrally synchronized EEG signals recorded during  {\it resting-state}. We provide technical background information about the study design and measurement structure in a web-based supplement. Here we investigate neuronal synchronicity in a group of typically developing (TD) children. We contrast group inference for the TD cohort against patterns of synchronicity in a cohort of children diagnosed with Autism  Spectrum Disorder (ASD) in \S~\ref{subsec: Results}. To our knowledge this is the first attempt at population level-inference for neuronal synchronicity in the setting of EEG studies.

\subsection{MIC Analysis of TD and ASD Children}
\label{subsec: Results}
Autism Spectrum Disorder (ASD) describes a neurodevelopmental condition, characterized by social communication deficits, presence of repetitive behaviors, and/or restricted interest. Clinical presentation is highly variable, with heterogeneity in relation to medical conditions, behavioral challenges, and degree of intellectual impairments  \citealt{parr2011early}).  Such behavioral and neurophysiological heterogeneity poses serious challenges to the study of the neurophysiological substrate. In this respect, resting-state EEG is a particularly advantageous, and therefore popular, brain imaging choice (\citealt{wang2013resting}). 

Here we perform a comparative study between age-matched TD and ASD cohorts, under the framework of Multilevel Integrative Clustering (MIC). The study includes 9 participants (29-60 months of age) from the TD group, and 10 participants (27-99 months of age) from the ASD group. During the experiment, EEG was recorded at 250Hz using 129 channel geodesic nets with Ag/AgCl electrodes. Recordings took place while participants watched videos of bubbles and other non-social images on a computer monitor for 2 to 6 minutes.

Starting with the TD cohort, our analysis follows the scheme detailed in \S~\ref{subsec: preprocessing} and considers epochs composed of  $\gamma=6$ contiguous 1024ms segments, allowing for a $\delta=0.5$ overlap between epochs. This choice was based on both substantive and empirical considerations. In particular, we consider a smoothing strategy that  guarantees good average adherence.   A sensitivity analysis to differential smoothing choices was carried out with respect to both the epoch length and the percent of overlap.  While details are reported in a supplementary document, we observe fairly robust results, with only small changes in estimation and selection of the number of clusters, echoing our findings in the simulation setting.

\begin{figure}
\begin{tabular}{c | c }
\footnotesize (1.a) \scriptsize ASD Subject 7   & \footnotesize  (1.b)  \scriptsize ASD Subject 9 \\
\includegraphics[width = 0.49\linewidth]{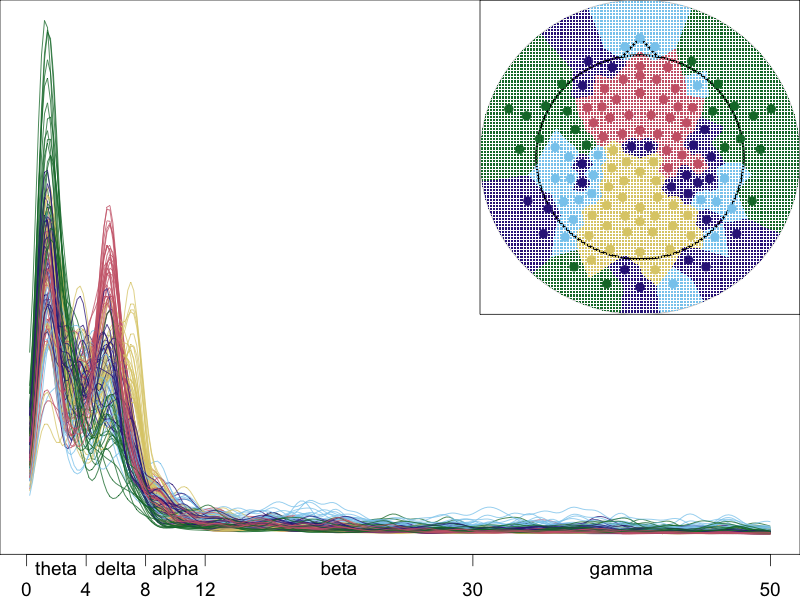}        &
\includegraphics[width = 0.49\linewidth]{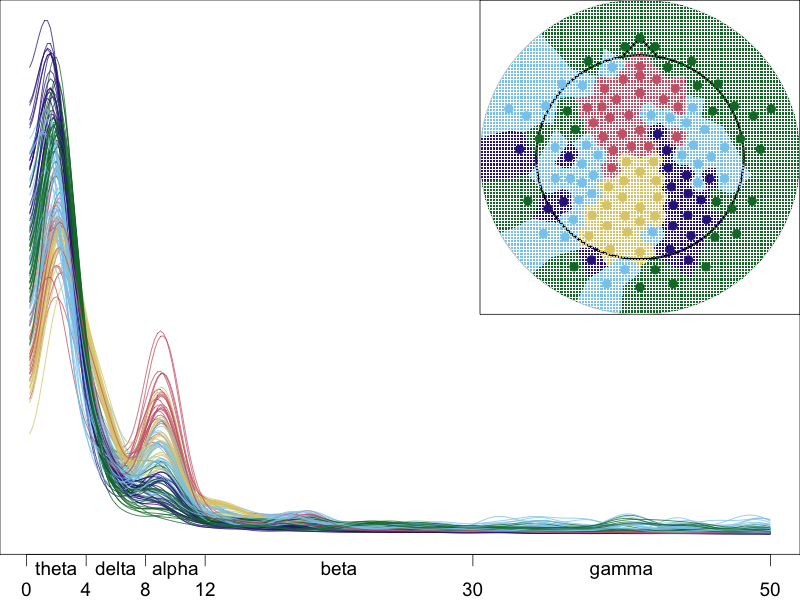}                    \\
\hline
&\\[-.1in]
\footnotesize (2.a) \scriptsize TD Subject 3  & \footnotesize  (2.b)  \scriptsize TD Subject 8 \\
\includegraphics[width = 0.49\linewidth]{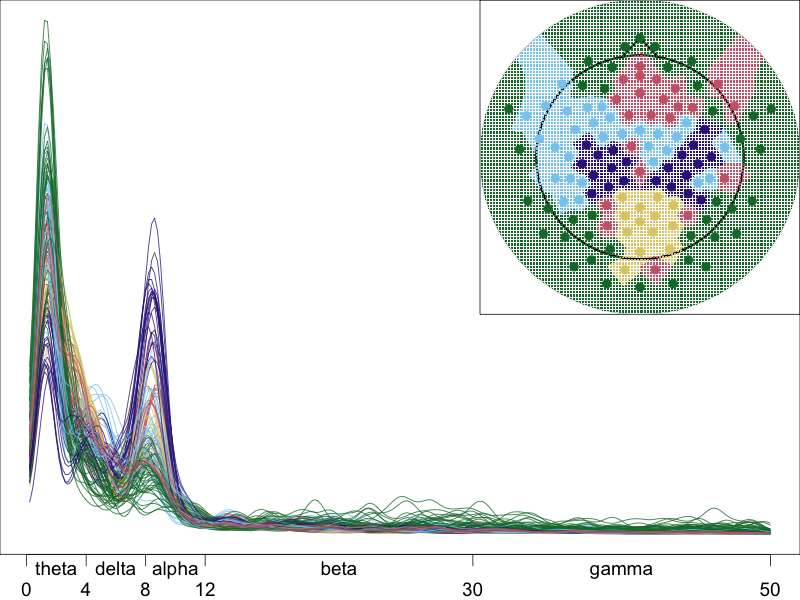}      &
\includegraphics[width = 0.49\linewidth]{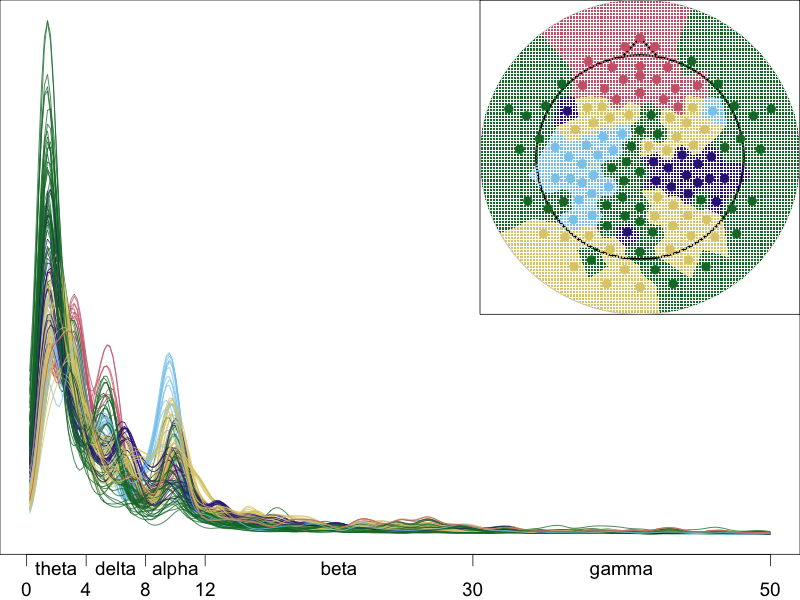}                 \\
\end{tabular}
\caption{\small {\bf Synchronicity and spectral features}:  For each cohort,  cluster configurations are depicted for two illustrative subjects. For each electrode, the estimated spectral density (normalized) is color coded by cluster membership. All plots refer to the epoch that is most coherent with subject-level clustering.}
\label{fig:ClustSpect}
\end{figure}

An illustration of how the proposed method clusters electrodes in relation to their spectral features is provided in Figure~\ref{fig:ClustSpect}. Here, we represent the epoch-level estimates of the spectral densities for each electrode, color-labeled by  inferred cluster membership. For each subject, we report the epoch of highest coherence with subject-level clustering. This simple illustration shows how, pooling information at the level of cluster labels can be achieved without requiring the spectral structure of  electrode-level time series to be aligned across subjects. We maintain that this feature is particularly appealing in resting-state neurocognitive settings, where complex and dynamic alignment issues may render extremely difficult any attempt at pooling EEG signals directly.

\begin{figure}
\begin{tabular}{c | c | c}
\footnotesize (1.a) \scriptsize ASD Population   & \footnotesize  (1.b)  \scriptsize ASD entropy & \footnotesize  (1.c) \scriptsize ASD cluster assignments\\
\includegraphics[width = 0.31\linewidth]{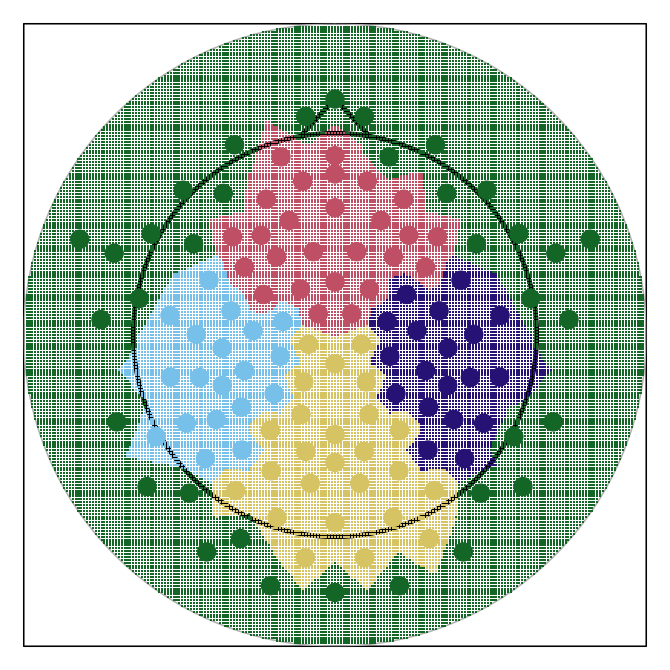}        &
\includegraphics[width = 0.32\linewidth]{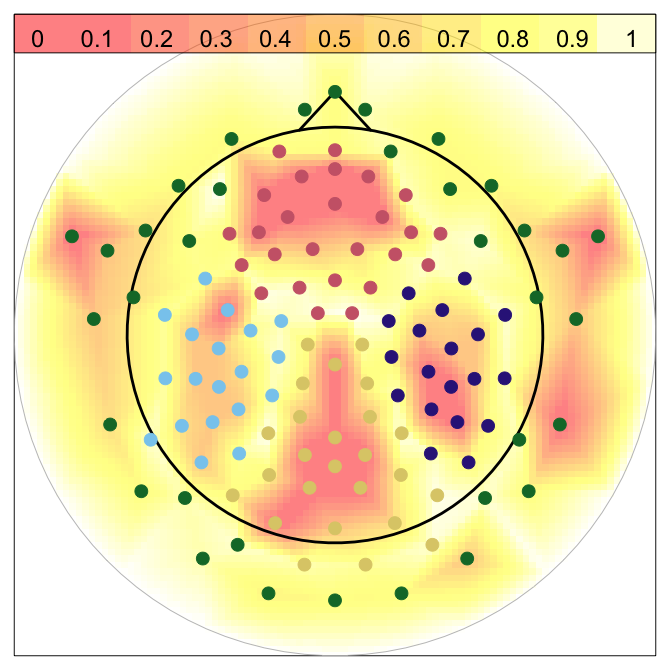}  &
\includegraphics[width = 0.31\linewidth]{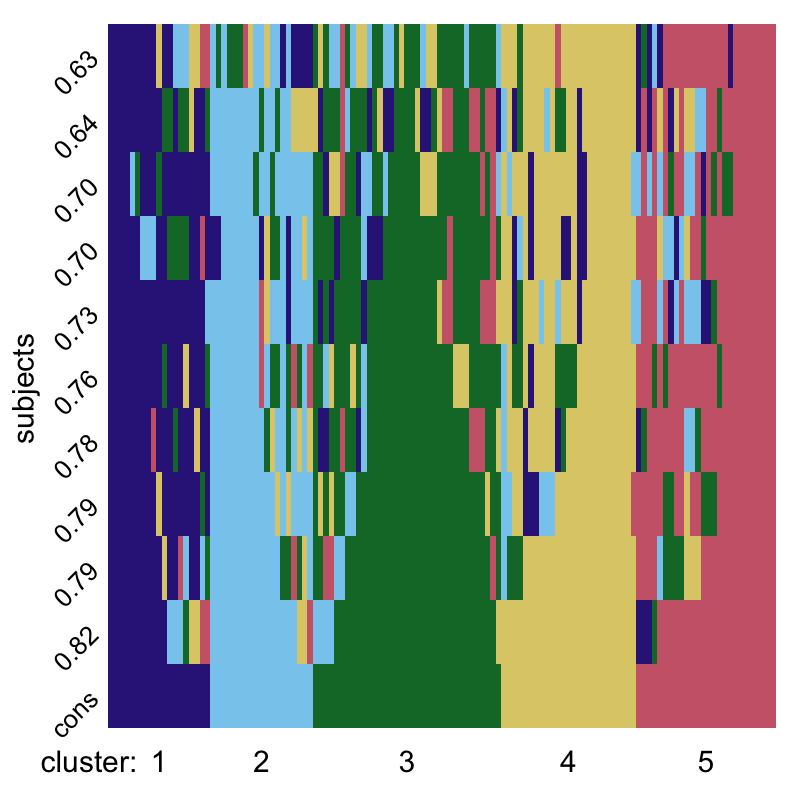}\\
\hline
&&\\[-.1in]
\footnotesize (2.a) \scriptsize TD Population   & \footnotesize  (2.b)  \scriptsize TD entropy & \footnotesize  (2.c) \scriptsize TD cluster assignments\\
\includegraphics[width = 0.31\linewidth]{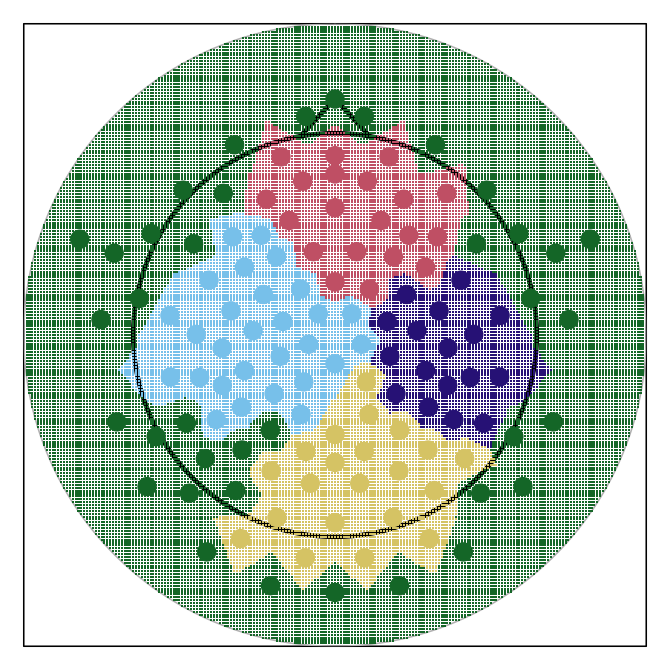}           &
\includegraphics[width = 0.32\linewidth]{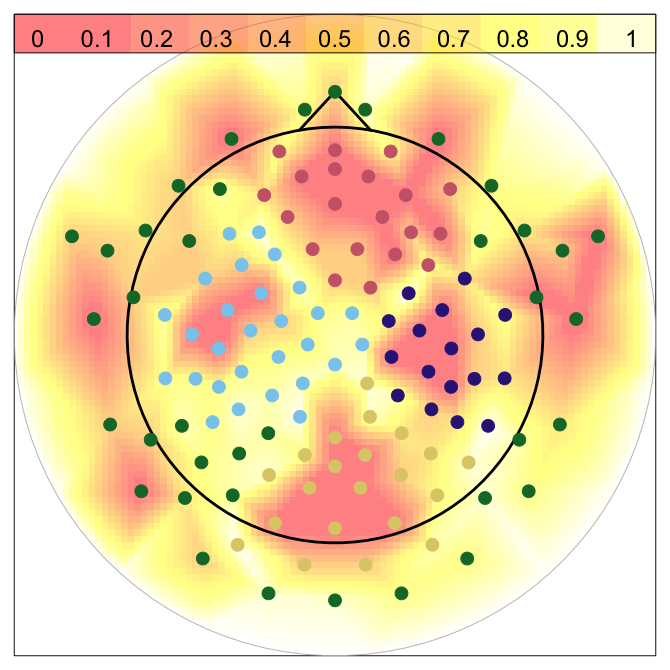}     &
\includegraphics[width = 0.31\linewidth]{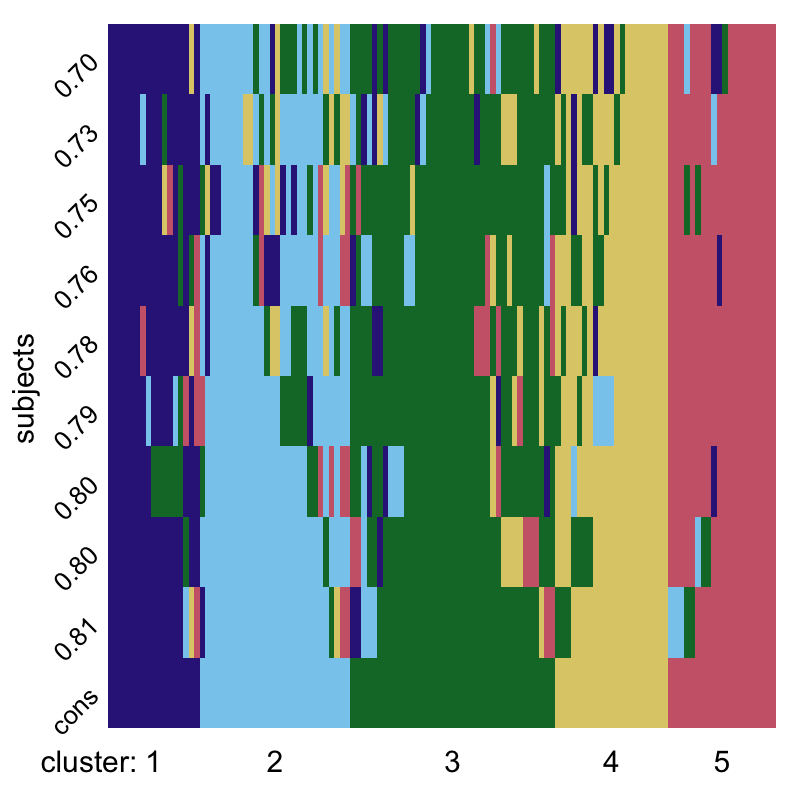}\\
\end{tabular}
\caption{\small {\bf Group contrasts, ASD (1) vs TD (2)}: 
(1.a) TD-cohort posterior least square synchronicity.
(1.b) TD-cohort normalized posterior entropy. 
(1.c) TD-cohort subject- and population-level cluster assignments. 
(2.a) ASD-cohort posterior least square synchronicity.
(2.b) ASD-cohort normalized posterior entropy.
(2.c) ASD-cohort subject- and population-level cluster assignments. 
In the (c) panels, we report consensus labels as the last row. Subject-level labels are reported in each row, together with  posterior median estimates of cluster adherence. }
\label{fig:ASDandTD}
\end{figure}

An informal comparison between TD and ASD groups is carried out in Figure~\ref{fig:ASDandTD}. For both cohorts,  we identify 5 spectrally synchronized areas, corresponding to the following cortical regions: frontal, left and right parieto-temporal, occipital, and peripheral, defined as a ring of outsidemost electrodes.  At the population level, the least square estimates of cortical clusters are remarkably similar between the two cohorts,  with the exception of an asymmetrical partition on  the occipital and parieto-temporal regions, where the left parieto-temporal cluster seems to be leaning towards the left hemisphere for ASD, but towards the right hemisphere for TD.

Further, we examine local and global sources of cluster variability in both groups.  At the electrode level, we report the entropy associated with posterior cluster label probabilities in Figure~\ref{fig:ASDandTD}: (1.b) for ASD and (2.b) for TD. Perfect partitions, e.g. an electrode assigned to cluster $k$ with probability one, yields 0 entropy, whereas uniform assignment probabilities yield entropy equal to 1. We observe that the mid-, right-frontal and mid posterior regions are the most stable regions for both groups.  Compared to the ASD group, the TD cohort exhibits more stable regions, for example, in the left-temporal (speech and language related), left-central, as well as some regions in the posterior and occipital areas of the cortex. The high entropy observed on the left-hemisphere among ASD children coincides with the abnormal left-hemispheric asymmetry findings in the literature on individuals with ASD (\citealt{stroganova2007abnormal}, \citealt{burnette2011anterior}). 

We gain more insight into the nature of variability of synchronized neuronal patterns by examining global sources of cluster variance at the subject-specific and population levels. In Figure~\ref{fig:ASDandTD}: (1.c) for ASD and (2.c), we report subject and population level cluster assignments for both TD and ASD cohorts. For each subject we also report the posterior median coherence estimate. We note how ASD children exhibit higher clustering  heterogeneity, with coherence estimates ranging from 0.63 to 0.82, compared to the TD cohort, with coherence estimates ranging from 0.70 to 0.81.  A similar conclusion is noted  in the higher entropy associated with ASD consensus estimates (1.b and 2.b).  This observation echoes some of our previous findings in EEG studies of implicit-learning in ASD and TD children (\citealt{Hasenstab:2015};  
\citealt{Hasenstab:2016a}; \citealt{Hasenstab:2016b}).

While formal covariate adjustments are outside the scope of this manuscript, we attempted a post-hock analysis aimed at explaining subject-level cluster variability using subjects age, ASD vs. TD cohort indicators,  and electrode-level band power estimates. Using cluster labels as a categorical outcome, we used random forests as a flexible tool to get a sense of variable importance in the classification of synchronous electrodes. We found the out-of-bag estimate of classification accuracy to be about 0.73, with subject's age explaining the largest mean decrease in accuracy and therefore being flagged as one of the most important predictors. None of the power bands had a specific predictive advantage in explaining subject-level cluster variability, confirming our intuition that, in the setting of resting state experiments, it may be inappropriate to pool subject-level spectral features directly, in order to infer connectivity. A less stringent model, like the one proposed in this manuscript, is therefore likely to be more robust in applications. More details for this analysis are included in the supplementary materials.

%
% Discussion -----------------------------------------------------------
%
\section{Discussion}
\label{sec:Discussion}

This paper proposes what to our knowledge is the first comprehensive statistical framework for population level inference of spectrally synchronized brain activity from a heterogeneous sample of EEG readings. A hierarchical model allows for the estimation of population level synchronicity patterns, with full consideration of intra- and inter-subjects variability. Crucially, information is borrowed at the latent level of cluster membership indicators. Dependent mixtures are based on a hierarchical Dirichlet prior, indexed by interpretable and informative parameters, which measure cluster adherence at all levels of the hierarchy.

Our approach melds non-parametric dimension reduction and fully model-based techniques through a graph-partitioning representation of clustering. This two-stage approach is likely to be useful in several experimental settings involving EEG measurements, where different scientific goals and different  data meta-processing concerns may require substantial subject-matter input in the definition of similarity between cortical regions. 

In our study we operate within the context of spectral synchronicity. It is however important to point out that alternative measures of neuronal affinity, for example partial correlation, coherence, and mutual information,  are also amenable to MIC analysis. In this sense, the proposed framework is quite general and can be adapted to handle alternative neuroimaging data platforms, such as functional Magnetic Resonance Imaging (fMRI). This consideration also applies, with possible minor adjustments, to the integration of multiple imaging modalities. This flexibility traces back to the hierarchical prior, which relates cluster labels rather than cluster-specific parameters (location and scale for example), so that complex data alignment issues are resolved within a higher level of modeling abstraction. Clearly, technical preprocessing pipelines may differ substantially between and within modalities. Therefore, important analytic details should be thoughtfully engineered in practice.

Our simulation results in \S~\ref{subsec:simout}, show that inference is robust to reasonable variants in the meta-processing strategies. In our experiments, simple information criteria like  BIC tend do do well in the selection of the number of clusters $K$, when combined with a search over the number of eigen-Laplacians $d$. Our model, of course, offers a very simple representation of  cluster variability within- and between-subjects. Therefore, modeling refinements are likely needed in applications where one can expect a strong dynamic evolution of synchronicity patterns; such as the setting
of stimulus-based EEG studies. 

Potentially useful extensions include a formal treatment of group comparison and covariate adjustments. In particular, predictors could, in principle, be introduced through cohesion functions as in \cite{Muller:2011}. We note, however, that the multilevel and dynamic structure of cluster configurations may require significant efforts to extend available covariate adjustment strategies in clustering.  Other options would include covariates through a regression on subject-level coherence parameters, which would perhaps lead to simpler and more interpretable models. 

\vskip.1in
\noindent A user-friendly implementation of the proposed method is available online as an R package at: https://github.com/Qian-Li/MIC2.

\bibliographystyle{ECA_jasa}
\bibliography{Connectivity}

\newpage

%
% FIGURES  AND TABLES  -----------------------------------------------------------------------------------------------------------
%
\graphicspath{{/JASA_plots/}}

\begin{table}
\small
\begin{center}
\begin{tabular}{r ccc c ccc}
\hline
\hline
           & \multicolumn{3}{l}{Stationary Setting} & &\multicolumn{3}{l}{Locally Stationary Setting}\\[.02in]  
SNR    & N=10  & N=20  & N=40 & & N=10 & N=20 & N=40 \\
\hline
10        & 0.812 &  0.868 & 0.903 & & 0.808 & 0.868 & 0.897 \\
5          & 0.810 &  0.867 & 0.898 & & 0.810 & 0.868 & 0.898 \\
1          & 0.810 &  0.869 & 0.900 & & 0.813 & 0.866 & 0.902 \\
\hline
\hline
\end{tabular}
\end{center}
\caption{\small Simulation study: Group-level clustering accuracy for varying sample size and signal to noise ratio.}
\label{tab:simulation}
\end{table}

%
% FIG 1 [Simulation Spectral Density]
%
\DeclareGraphicsExtensions{.png}

%
% FIG 2 [Simulation Convergence and Accuracy]
%
%

%
% FIG 3
%
%% TD vs ASD plots

%
% FIG 4
%

\end{document}